%% file: ms.tex
\documentclass[notitlepage,apj,numberedappendix]{emulateapj}

\usepackage{color}
\usepackage{natbib}
\usepackage{graphicx}
\usepackage{bibentry}
\usepackage{enumitem}

\newcommand{\simgt}{\,\hbox{\lower0.6ex\hbox{$\sim$}\llap{\raise0.6ex\hbox{$>$}}}\,}
\newcommand{\simlt}{\,\hbox{\lower0.6ex\hbox{$\sim$}\llap{\raise0.6ex\hbox{$<$}}}\,}

\begin{document}

\title{Testing Emergent Gravity with Optical, X-ray, and Weak Lensing Measurements in Massive, Relaxed Galaxy Clusters}
\author{J. A. ZuHone}
\affiliation{Harvard-Smithsonian Center for Astrophysics, 60 Garden St., Cambridge, MA 02138, USA}
\and
\author{J. Sims}
\affiliation{Covenant Christian Academy, 83 Pine St., Peabody, MA 01960, USA}

\begin{abstract}
We test the predictions of Emergent Gravity using matter densities of relaxed, massive clusters of galaxies using observations from optical and X-ray wavebands. We improve upon previous work in this area by including the baryon mass contribution of the brightest cluster galaxy in each system, in addition to total mass profiles from gravitational lensing and mass profiles of the X-ray emitting gas from \textit{Chandra}. We use this data in the context of Emergent Gravity to predict the ``apparent'' dark matter distribution from the observed baryon distribution, and vice-versa. We find that although the inclusion of the brightest cluster galaxy in the analysis improves the agreement with observations in the inner regions of the clusters ($r \simlt 10-30$~kpc), at larger radii ($r \sim 100-200$~kpc) the Emergent Gravity predictions for mass profiles and baryon mass fractions are discrepant with observations by a factor of up to $\sim2-6$, though the agreement improves at radii near $r_{500}$. At least in its current form, Emergent Gravity does not appear to reproduce the observed characteristics of relaxed galaxy clusters as well as cold dark matter models.
\end{abstract}

\section{Introduction}\label{sec:intro}

Astronomical observations at galactic and cosmological scales indicate that roughly 95\% of the total energy budget in the universe is in an unknown form. The total matter density in the universe is inferred to be roughly 5$\times$ that of the baryonic matter density based on observations of rotation curves of disk galaxies, stellar velocity dispersions in elliptical galaxies, the motions of galaxies and temperatures of hot gas in galaxy clusters, gravitational lensing of galaxies and clusters, and analysis of the structure of the cosmic microwave background. The consensus in the astrophysical community is that this deficit is made up for by gravitating matter which does not interact electromagnetically, hence the name ``dark matter'' (hereafter DM). Nearly 70\% of the energy budget of the universe is made of an unknown substance dubbed ``dark energy'' which is consistent with Einstein's cosmological constant $\Lambda$ within current measurement limits \citep[see, for example, the recent constraints from][]{planck2018}.

Since no candidate particles for DM have yet been unambiguously identified either in particle physics experiments or astronomical observations, other authors propose that the observations mentioned above provide evidence for a departure from Newton's (and Einstein's) law of gravity on galactic and cosmological length scales. The most famous of these theories is of course Modified Newtonian Dynamics (MOND), proposed originally by \citet{milgrom1983}.

Recently, \citet{verlinde2017} proposed a modified theory of gravity based on the emergence of spacetime and gravity from the entanglement structure of an underlying microscopic theory, known as ``Emergent Gravity'' (hereafter EG). According to EG, the microscopic states in de Sitter space form an elastic response to matter at length scales smaller than the Hubble radius. In the classical limit, this response creates an additional gravitational force over and above that from baryons, which potentially explains the phenomena mentioned above without the need for positing an additional, non-baryonic form of matter. 

Though the evidence for the astrophysical consensus on DM is abundant, alternative theories such as that of \citet{verlinde2017} are worth examining in the light of observations. This task has already been carried out by other authors, who have used the baryon masses observed in astrophysical systems to predict the expected additional acceleration predicted by EG. The results are so far mixed. \citet{brouwer2017} and \citet{tortora2018} found agreement with the predictions of EG using galaxy-galaxy lensing and dynamical mass estimates, respectively. \citet{lelli2017} showed a disagreement between EG and the radial acceleration relation in observations of disc galaxies, though \citet{hossenfelder2018} claimed an agreement between this relation and their covariant version of EG derived in \citet{hossenfelder2017}. \citet{dt2018} showed marginal agreement of velocity dispersion profiles in the Milky Way's dwarf spheroidals with EG, but \citet{pardo2017} found for a sample of isolated dwarf galaxies that EG underpredicted their maximum velocities for more massive systems. 

A few other works have compared the predictions of EG to observations of clusters of galaxies. Galaxy clusters are the largest bound objects in the universe to have formed via the process of hierarchical structure formation, hence they contain a fair representation of the different kinds of matter in the universe. This makes them excellent probes of the properties of the ``dark sector.'' In the case of confronting the predictions of EG with observations of clusters of galaxies, here too the results are mixed. \citet{ettori2017} and \citet{ettori2019} used clusters from their ``{\it XMM-Newton} Cluster Outskirt Programme'' \citep{eckert2017} to test the predictions of EG out to large radius. Using two relaxed clusters (A2142 and A2319), they found that EG underpredicts the apparent DM mass by a factor of up to a few in the innermost regions of the cluster, but the prediction improved as a function of increasing radius until $\sim{r}_{500}$, where the EG prediction and the DM mass estimated from hydrostatic equilibrium matched. Similar trends were noticed by other authors for a number of other systems \citep{hodson2017,halenka2018,tam2019}.

\input{clusters.tex}

In this work, we again test the predictions of EG against observations of matter densities of galaxy clusters, but we build upon the previous approaches to this problem in two ways. The first is that we extend the analysis of the apparent DM distribution predicted by EG to the core region by including the contribution to the baryon mass of the brightest cluster galaxy (BCG), which is the dominant component in the center of the cluster. The effect of the BCG was not considered in the previous studies mentioned above. The BCG will dominate over both the hot plasma from clusters (the intracluster medium, or ICM) and the stellar contribution from other galaxies and intracluster light within the inner $\sim30$~kpc of the cluster center. The second extension of our work over the previous studies is that we also examine the predictions made by the covariant version of EG from \citet{hossenfelder2017}. Aside from exploring this alternative to the standard CDM model, we assume a spatially flat $\Lambda$-dominated cosmology with $h = 0.7$, $\Omega_{\rm m}$ = 0.3, and $\Omega_{\Lambda}$ = 0.7. 

\section{Data}

\begin{figure}
\centering
\includegraphics[width=0.45\textwidth]{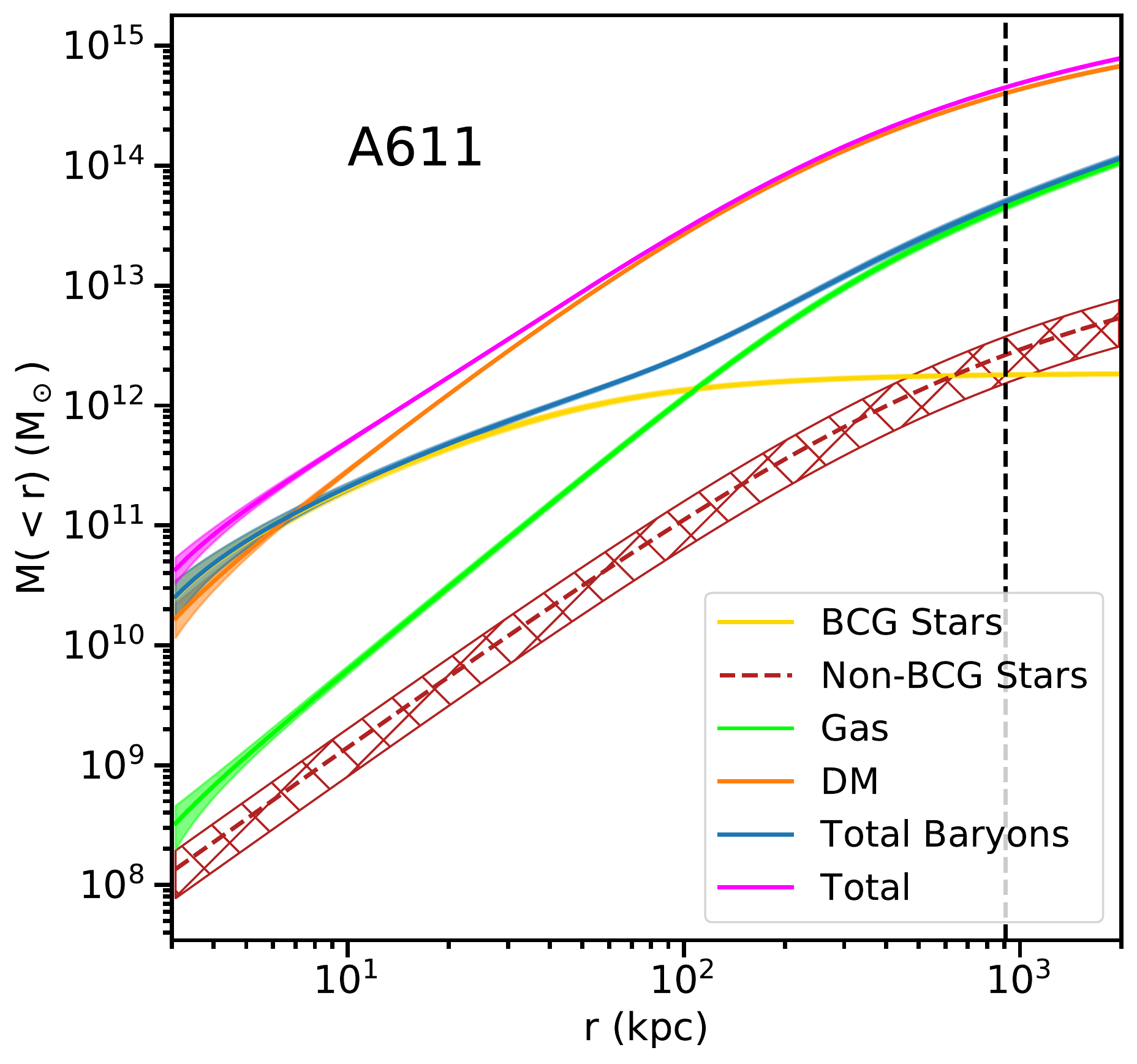}
\caption{Enclosed mass profiles for A611 for all of the contributions considered in our calculations. Solid lines indicate mean values and the shaded regions surrounding them indicate 1-$\sigma$ errors. The non-BCG stellar contribution is shown as a dashed line with a hatched error region to indicate it is an estimate and does not come from direct observations of the cluster. The vertical dashed line indicates the location of $r_{500}$.\label{fig:A611_example_profiles}}
\end{figure}

The four clusters we examine in this work are A611, A963, A383, and A2390. These are massive ($M_{200} \sim 10^{15}~M_\odot$), fairly nearby ($z \sim 0.2-0.3$), relatively relaxed systems, with massive BCGs in their centers \citep{newman2013a}. The basic properties of these systems are listed in Table \ref{tab:clusters}. In what follows, we provide an accounting of the observed mass components from these clusters we included in our analysis.  

For the total mass and BCG density of the clusters, we use the radial profiles derived from strong and weak lensing and resolved stellar kinematic measurements first reported in \citet{newman2013a} and analyzed further in \citet{newman2013b}. The BCG profiles can be modeled by a ``dPIE'' profile \citep{eliasdottir2007}:
\begin{equation}\label{eqn:dpie}
\rho_{\rm BCG}(r) = \frac{\rho_0}{(1+r^2/r_{\rm core}^2)(1+r^2/r_{\rm cut}^2)}
\end{equation}
where $\rho_0$ is the central density and $r_{\rm core}$ and $r_{\rm cut}$ are core and cutoff radii for the profile, respectively. \citet{newman2013a,newman2013b} noted that the total mass density is dominated by the BCG in the innermost core region $r \simlt 10-20$~kpc. They showed the rest of the cluster is dominated by DM, and is well-represented by a profile of the form \citep[][]{zhao1996}:
\begin{equation}
\rho_{\rm DM+gas}(r) = \frac{\rho_s}{(r/r_s)^\beta(1+r/r_s)^{3-\beta}}
\end{equation}
where $\rho_s$ is the scale density, $r_s$ is the scale radius, and the inner slope $\beta$ for the DM is typically shallower than $\beta = 1$ for these systems, which is the value of $\beta$ corresponding to the \citet[][hereafter NFW]{nfw97} profile. They also showed that the total mass profiles are well-fit by the NFW profile itself. The total mass and BCG density profiles and their uncertainties \citep[including all sources of statistical and systematic error included in][]{newman2013a,newman2013b} were provided by A. Newman via private communication. 

The dominant baryonic component by mass in galaxy clusters is the hot X-ray emitting plasma. To include this component in our analysis, we use gas mass density profiles from \citet{martino2014}, who fit the following equation from \citet{vikhlinin2006} for the emission measure to data from \textit{Chandra}, including for our four clusters: 
\begin{equation}\label{eqn:nenp}
n_{\rm p}n_{\rm e}(r)=\frac{n^2_{\rm 0}(r/r_{\rm c})^{-\alpha}}{[1+(r/r_{\rm c})^2]^{3\beta-\alpha/2}}+\frac{n^2_{\rm 02}}{[1+(r/r_{\rm c2})^2]^{3\beta_{\rm 2}}}.
\end{equation}
where $\alpha, \beta$, and $\beta_{\rm 2}$ are slope parameters, $n_{\rm 0}$ and $n_{\rm 02}$ are density normalization parameters, and $r_{\rm c}$ and $r_{\rm c2}$ are core radii. From this equation a gas density profile can be defined via $\rho_g = 1.252m_p\sqrt{n_pn_e}$. For this paper, the values of the parameters and their error bars were provided by P. Mazzotta via private communication. The available X-ray data does not constrain the emission measure profiles of our four clusters well beyond $r_{500}$, hence we do not attempt to calculate the effects of EG far beyond this radius in this work. \citet{martino2014} also calculated total mass profiles from the X-ray data under the assumptions of spherical symmetry and hydrostatic equilibrium; we have compared these profiles to the total mass profiles derived from weak lensing and found that they agree to within $\sim 20\%$ within $r_{500}$, expected for relatively relaxed clusters. All of the data used in this work assumed the same cosmology as we noted above in Section \ref{sec:intro}.

\begin{figure*}
\centering
\includegraphics[width=0.95\textwidth]{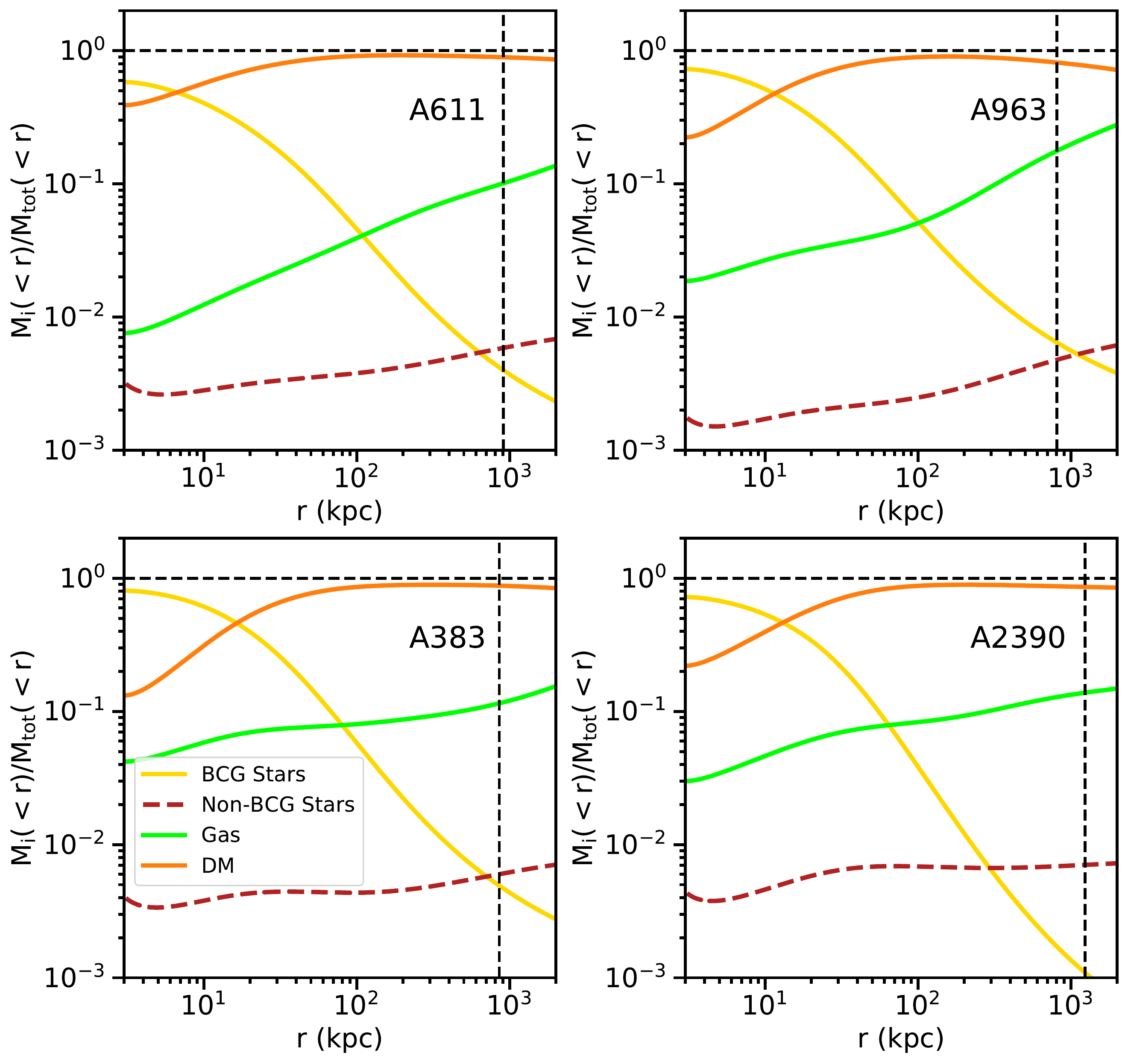}
\caption{Enclosed mass fraction profiles $M_i/M_{\rm tot}$ of the four major components by mass in our four clusters as a function of radius (shown without errors for clarity), showing the similarity between the different systems. The vertical dashed lines mark the position of $r_{500}$ for every system.\label{fig:all_mass_fractions}}
\end{figure*}

The only remaining mass component to take account of is the stellar component associated with the other member galaxies of the cluster. This component should be a small fraction of the total matter density. To represent it for each cluster, we follow the approach of \citet{ettori2017} and represent the satellite galaxy density profile using an NFW profile with a concentration parameter of $c_g = 2.9$ from the results of \citet{lin2004}. We normalize this profile for each cluster using the relationship between total cluster mass and satellite galaxy mass found by \citet{kravtsov2018}. We use Monte Carlo realizations to determine the error bars on the satellite galaxy mass profile using the error estimates and intrinsic scatter from \citet{lin2004} and \citet{kravtsov2018}. We have performed all of our calculations in this work with and without the contribution from this satellite galaxy component and found that it does not affect our conclusions.

Figure \ref{fig:A611_example_profiles} shows the radial enclosed mass profiles of all of these mass components for A611 as an example. The stellar component not associated with the BCG is shown as a dashed line with a hatched error region to indicate that this profile does not come from direct observations of the cluster but is only an estimate. The BCG contribution to the enclosed baryon mass is dominant over the contribution from the hot gas until roughly $r \sim 100$~kpc in this system, and is comparable to the contribution from the DM within the inner $\sim$10~kpc. Thus, it is crucial to include this component in our analyses.  

Figure \ref{fig:all_mass_fractions} shows the enclosed mass fraction for the four major components of each cluster for all four systems considered in this work. The basic structure of all four relaxed clusters is very similar. In the inner regions ($r \simlt 10-30$~kpc), the enclosed mass is dominated by the BCG, but at larger radii the mass contribution from the DM becomes the largest overall, whereas the gas becomes the dominant form of baryonic matter.

\section{Methods}

Throughout this work, we will follow the previous authors and cast the predictions of EG in the form of an ``apparent'' DM mass, to enable a intuitive comparison with the standard CDM theory. Here we will only provide a short description of the origin of the EG calculation of this apparent DM mass, for more extended discussions see \citet{verlinde2017} and \citet{tortora2018}. In the EG framework, the quantum entanglement entropy in a de Sitter spacetime adds a volume law contribution arising from dark energy in addition to the area law contribution from ordinary matter which is determined from the analysis of quantum fields in GR. In regions of large matter density, this dark energy entropy is displaced by baryonic matter and in the GR weak-field limit the Newtonian gravitational force law applies. 
\citet{verlinde2017} provided an expression relating the apparent DM mass to the observed baryon mass under the assumption of a static and spherically symmetric matter distribution:
\begin{equation}\label{eqn:eg_dm}
M_{\rm B}(<r) = \frac{6}{a_0r}\int_0^r\frac{GM^2_{\rm DM,EG}(<r')}{r'^2}dr'.
\end{equation}
where $M_{\rm B}$ and $M_{\rm DM,EG}$ are the enclosed baryonic and apparent DM mass as a function of radius, respectively, and $a_0 = cH_0$ is the ``acceleration scale''.

Equation \ref{eqn:eg_dm} postulates a relationship between the baryonic mass distribution in a static, spherically-symmetric system and the apparent DM distribution. Given an observation of one quantity, a prediction for the distribution of the other may be computed and compared to its observed distribution. If the EG scenario is correct, the observed mass profiles for both components should self-consistently transform into each other using Equation \ref{eqn:eg_dm}. This is in contrast to the DM model, where there is no \textit{a priori} reason for the DM and baryonic profiles to have a one-to-one relationship. 

Thus, we will carry out two analyses on our four clusters. First, we will predict an apparent DM mass profile from the observed baryonic mass profile, and compare the results to DM mass profiles derived from weak lensing data. The apparent DM mass profile can be derived from the observed baryonic mass profile by differentiating Equation \ref{eqn:eg_dm} with respect to the radius $r$, yielding \citep{ettori2017}:
\begin{equation}\label{eqn:eg_dm_reversed}
M^2_{\rm DM,EG}(<r) = \frac{a_0}{6G}r^2[M_{\rm B}(<r)+4\pi{r^3}\rho_{\rm B}(r)].
\end{equation}
We call this method ``EG1''.

\citet{hossenfelder2017} derived a generally covariant version of EG (CEG), providing a Lagrangian and deriving equations of motion. From Equation 10 of \citet{hossenfelder2018}, we can derive the apparent DM mass as a function of radius for a generic spherically symmetric system in the CEG framework, casting it into a similar form as Equation \ref{eqn:eg_dm_reversed}:
\begin{equation}\label{eqn:ceg_dm_reversed}
M^2_{\rm DM,CEG}(<r) = \frac{a_0}{6G}r^2M_{\rm B}(<r),
\end{equation}
which is very similar to Equation \ref{eqn:eg_dm_reversed} above but lacks the term depending explicitly on the baryon density. It should be noted that \citet{shen2018} points out that the derivation of Equation \ref{eqn:eg_dm} in \citep[][]{verlinde2017} contains an implicit assumption that the gravitational potential can be represented by a point mass, which is not true for extended mass distributions such as galaxy clusters. Assuming a general spherically symmetric mass distribution, \citet{shen2018} arrives at the same formula above from \citet{hossenfelder2017}. We will also compare the predicted apparent DM using this method, called ``CEG1''. 

The second method is to take the total mass profile $M_{\rm tot}$ from the same weak lensing data and, assuming the EG formalism, derive the baryonic mass profile which would produce it and compare it to the observed baryon profile. In this case, we assume that $M_{\rm tot} = M_{\rm DM,EG}+M_{\rm B}$ and derive a differential equation for $M_{\rm B}$ from Equation \ref{eqn:eg_dm}:
\begin{equation}\label{eqn:eg_dm_ode}
\frac{dM_{\rm B}}{dr}+\frac{M_{\rm B}}{r}-\frac{6G}{a_0}\frac{(M_{\rm tot}-M_{\rm B})^2}{r^3} = 0.
\end{equation}
which may be solved using normal ODE methods. We call this method ``EG2''. Note that this method does not incorporate any information about the baryon profile itself inferred from either X-ray (the ICM contribution) or optical (the stellar contribution) observations. We will also substitute $M_{\rm DM,CEG} = M_{\rm tot} - M_{\rm B}$ into Equation \ref{eqn:ceg_dm_reversed} above and solve for $M_{\rm B}$ assuming CEG, calling this method ``CEG2''. 
In both cases, we assume (as did the previous studies mentioned in Section \ref{sec:intro}) that $M_{\rm DM,EG} + M_{\rm B}$ produces the same weak-lensing signal as $M_{\rm DM} + M_{\rm B}$ for DM in GR. In the Newtonian limit, this implies that the gravitational acceleration in both cases is also equivalent. Uncertainties on physical quantities are propagated through our calculations by generating 3000 Monte-Carlo realizations of each input profile. 

\section{Results}

\subsection{Method EG1: Deriving Apparent DM Profiles from Baryon Profiles}\label{sec:EG1}

\begin{figure*}
\centering
\includegraphics[width=0.97\textwidth]{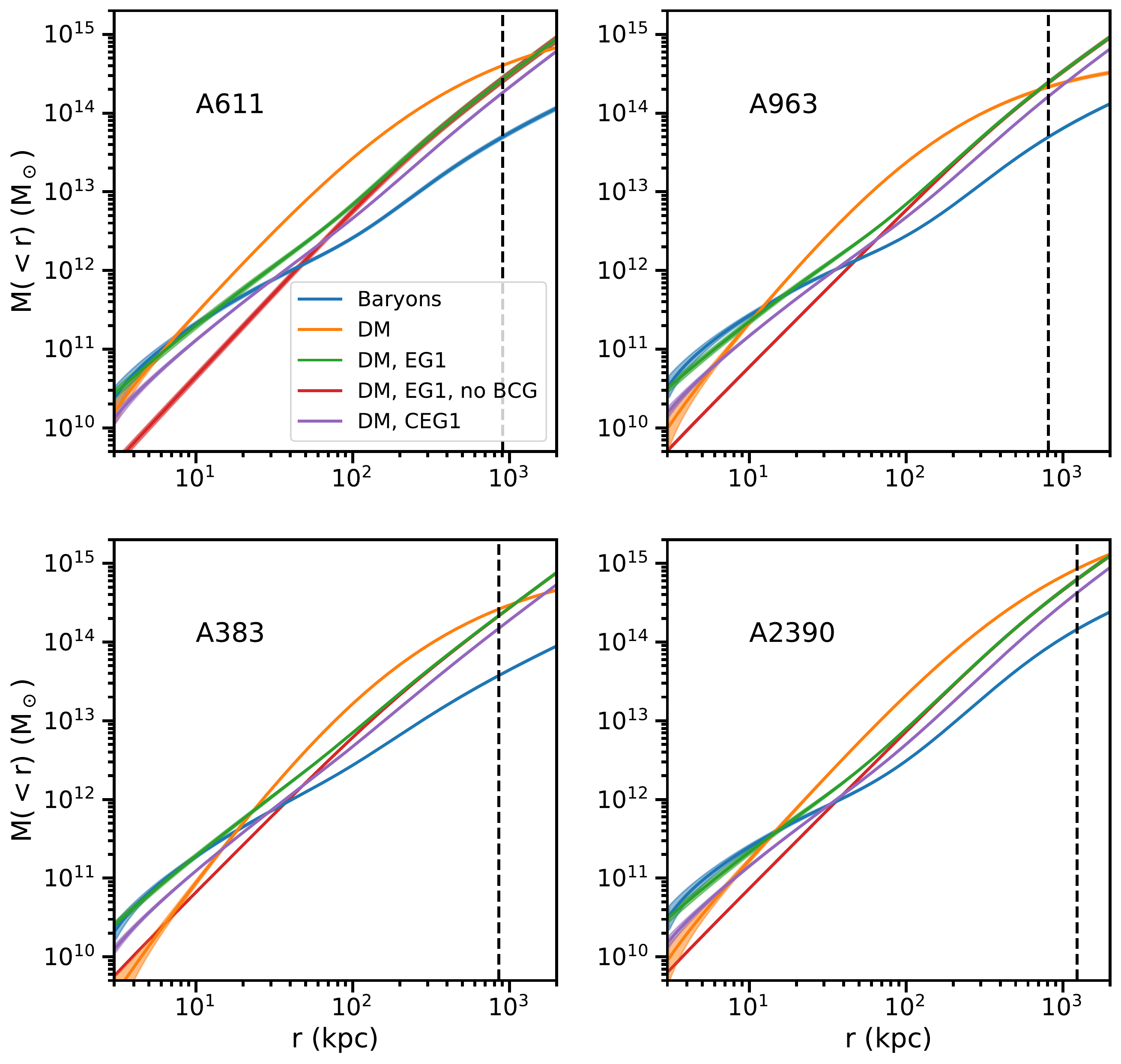}
\caption{Apparent DM mass profile derived from the EG formalism compared to that inferred from observations, using method EG1 and the observed baryon distribution, for all four clusters considered in this work. We consider cases where the BCG is included and where it is not. 1-$\sigma$ errors are indicated using the shaded bands. The location of $r_{500}$ for each cluster is indicated by the vertical dashed line.\label{fig:eg1_mass}}
\end{figure*}

\begin{figure*}
\centering
\includegraphics[width=0.97\textwidth]{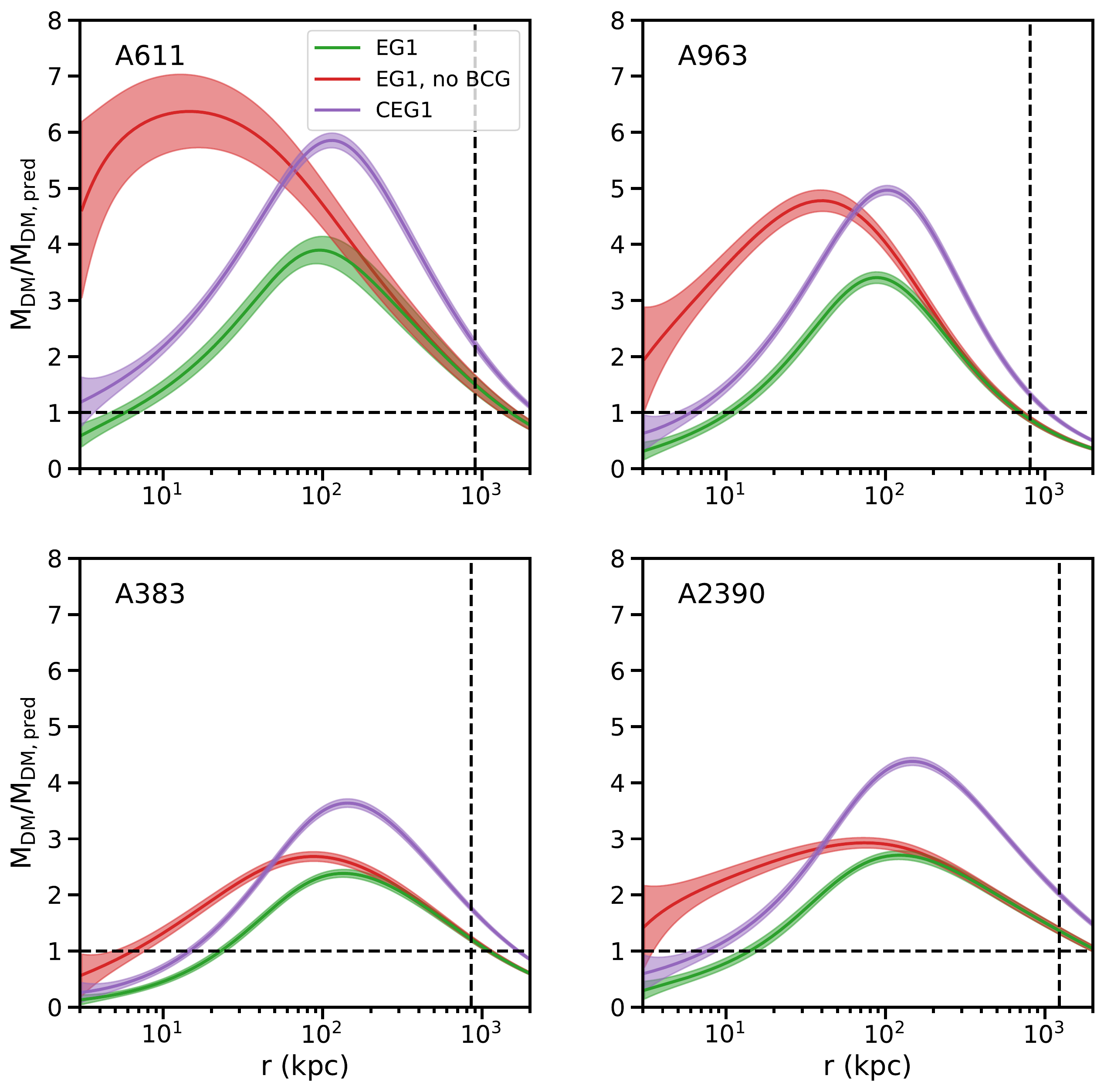}
\caption{Ratio of the observed DM mass the apparent DM mass derived from the EG formalism using method EG1 and the observed baryon distribution, for all four clusters considered in this work. We consider cases where the BCG is included and where it is not. 1-$\sigma$ errors are indicated using the shaded bands. The location of $r_{500}$ for each cluster is indicated by the vertical dashed line.\label{fig:eg1_mass_ratio}}
\end{figure*}

\begin{figure*}
\centering
\includegraphics[width=0.97\textwidth]{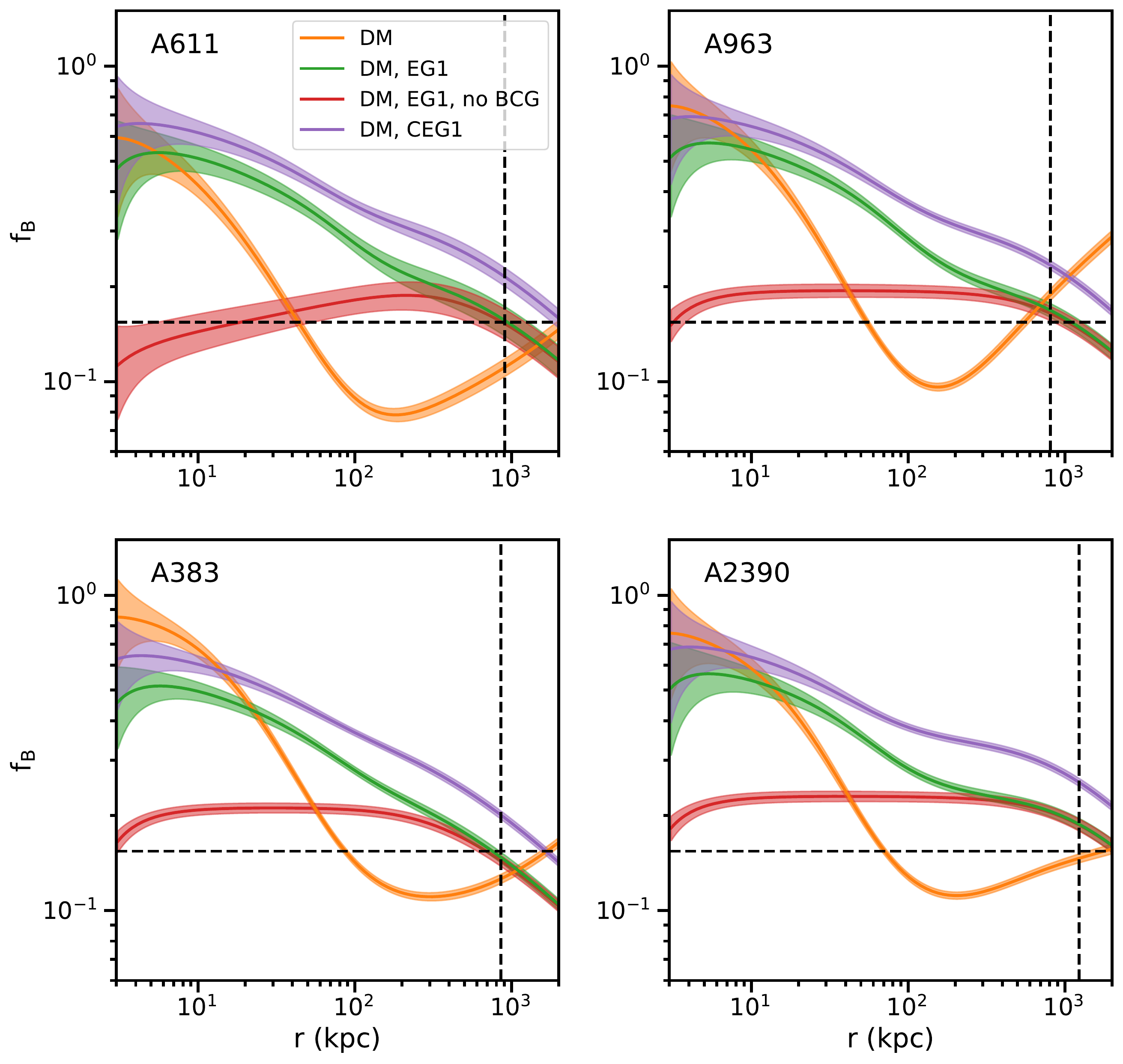}
\caption{Observed baryon mass fractions compared to those derived from the EG formalism using method EG1 for all four clusters considered in this work. 1-$\sigma$ errors are indicated using the shaded bands. The location of $r_{500}$ for each cluster is indicated by the vertical dashed line, and the cosmic baryon fraction $\Omega_{\rm b}/\Omega_{\rm m}$ \citep[][]{planck2018} is marked by the horizontal dashed line.\label{fig:eg1_b_fraction}}
\end{figure*}

\begin{figure*}
\centering
\includegraphics[width=0.97\textwidth]{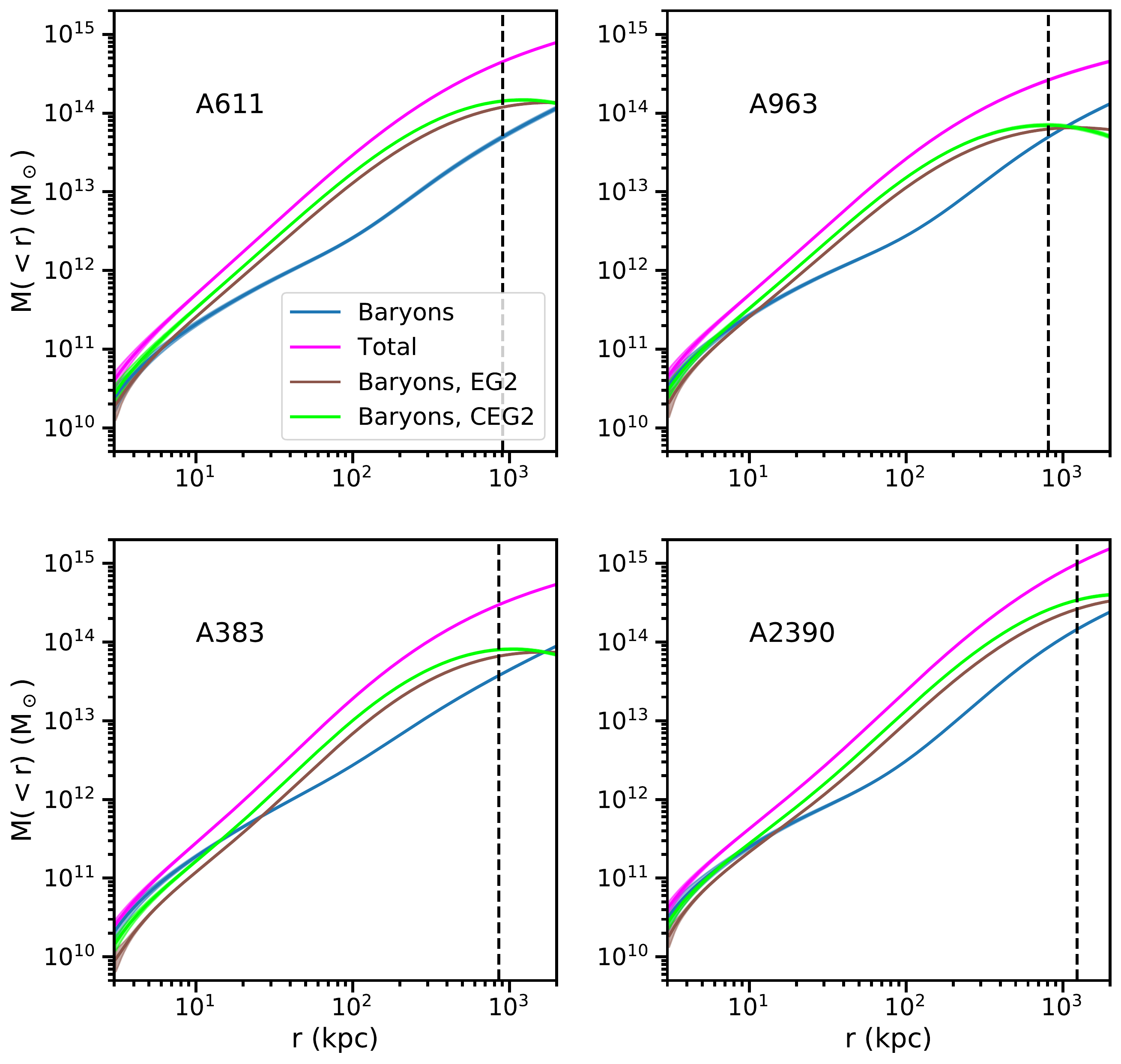}
\caption{The baryon mass profile derived from the EG formalism compared to that inferred from observations, using method EG2 and the observed total mass distribution, for all four clusters considered in this work. 1-$\sigma$ errors are indicated using the shaded bands. The location of $r_{500}$ for each cluster is indicated by the vertical dashed line.\label{fig:eg2_mass}}
\end{figure*}

\begin{figure*}
\centering
\includegraphics[width=0.97\textwidth]{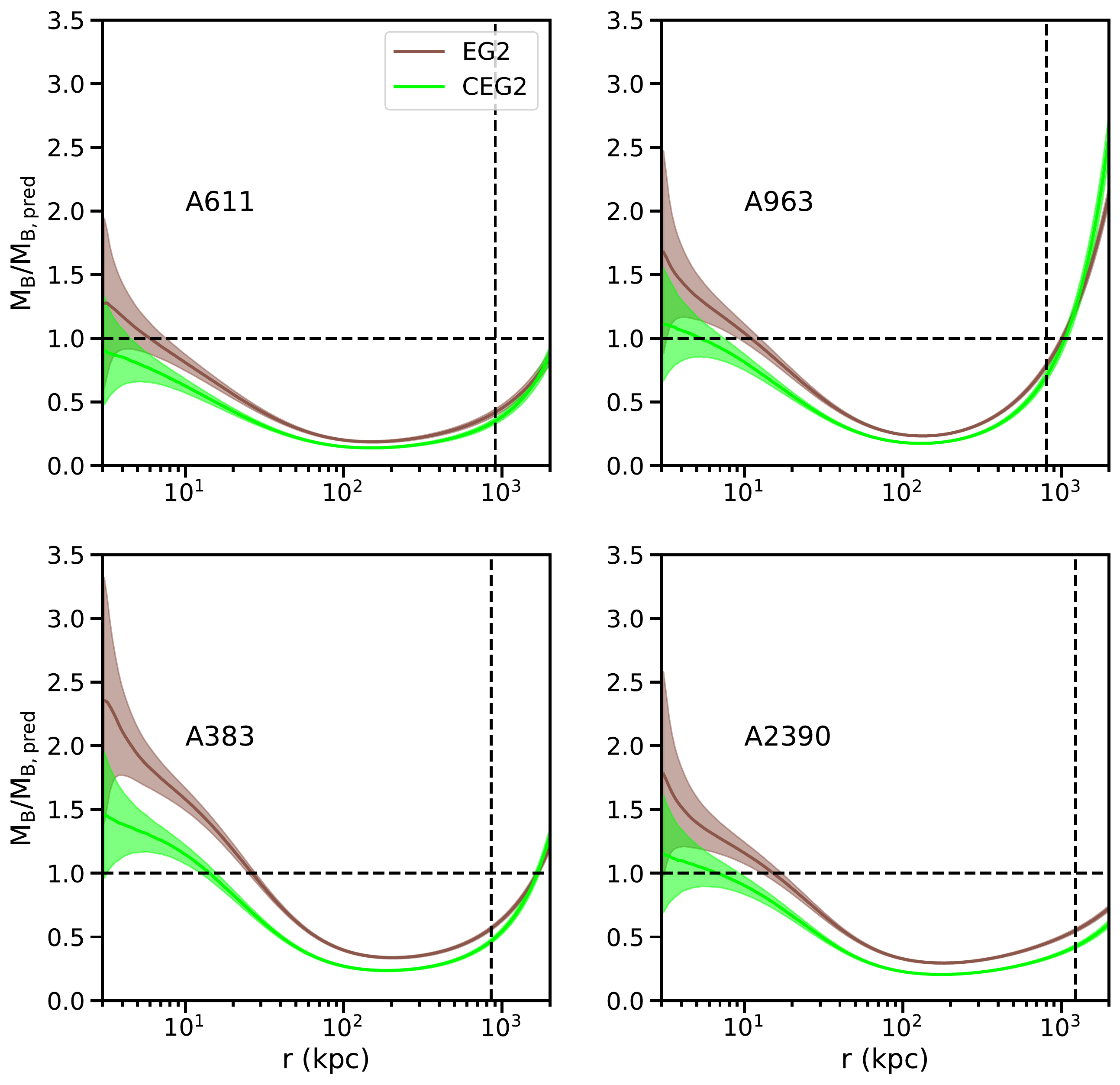}
\caption{Ratio of the observed baryon mass to the baryon mass derived from the EG formalism using method EG2 and the observed total mass distribution, for all four clusters considered in this work. 1-$\sigma$ errors are indicated using the shaded bands. The location of $r_{500}$ for each cluster is indicated by the vertical dashed line.\label{fig:eg2_mass_ratio}}
\end{figure*}

\begin{figure*}
\centering
\includegraphics[width=0.97\textwidth]{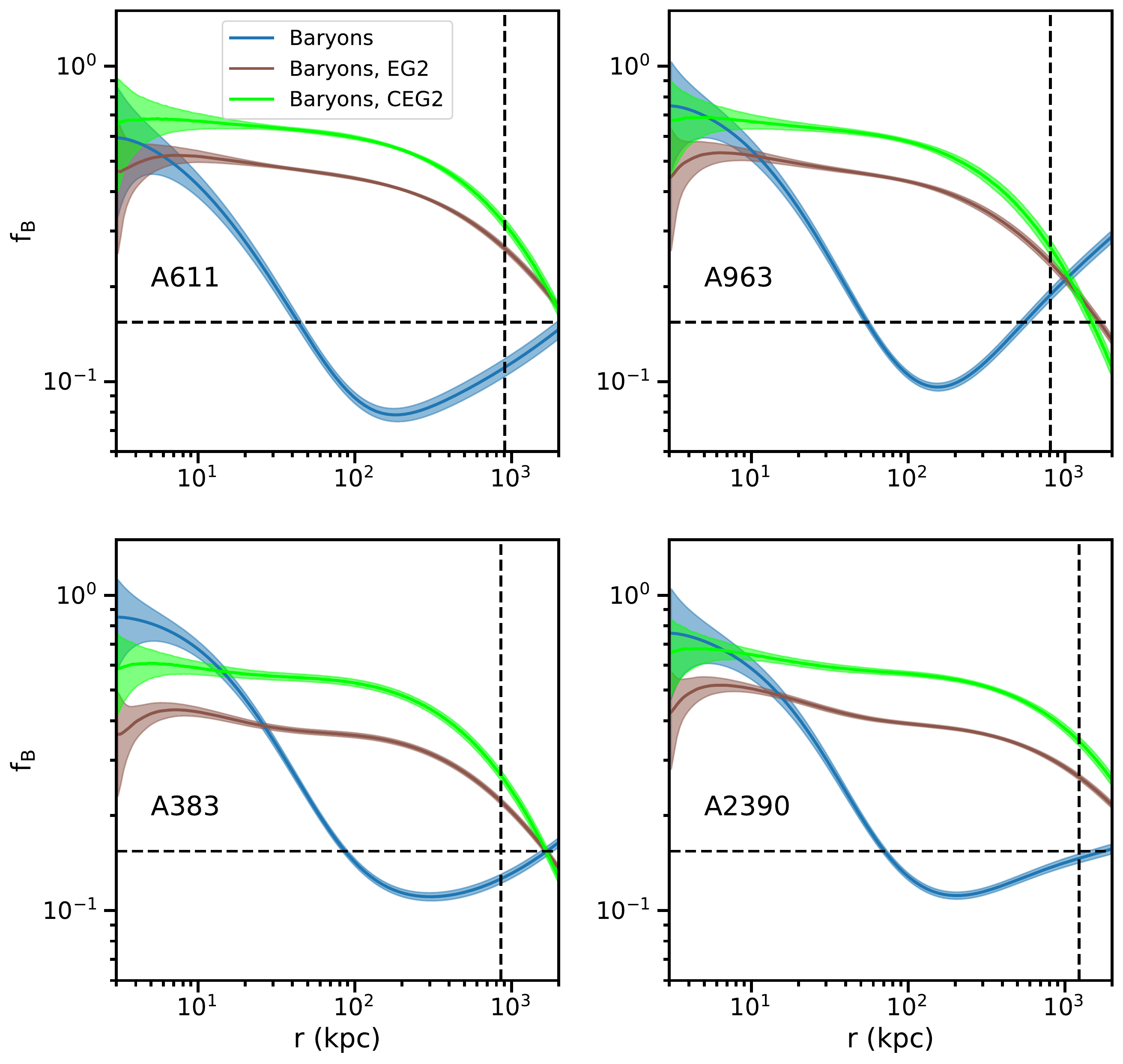}
\caption{Baryon mass fractions derived from the standard GR+DM model and the EG formalism using method EG2 and the observed total mass distribution, for all four clusters considered in this work. 1-$\sigma$ errors are indicated using the shaded bands. The location of $r_{500}$ for each cluster is indicated by the vertical dashed line, and the cosmic baryon fraction $\Omega_{\rm b}/\Omega_{\rm tot}$ \citep[][]{planck2018} is marked by the horizontal dashed line.\label{fig:eg2_b_fraction}}
\end{figure*}
We used Equation \ref{eqn:eg_dm_reversed} to derive the apparent DM profiles predicted by EG shown in Figure \ref{fig:eg1_mass}. Comparing the predicted apparent DM contribution from EG (green curves) to that inferred from gravitational lensing (orange curves), we see that for most of the radial range of the clusters outside of the central regions that EG underpredicts the observed DM. However, in the inner regions of the clusters where the BCG dominates, EG {\it overpredicts} the observed DM. In all four clusters, the apparent DM mass predicted by EG traces very closely the baryonic component dominated by the BCG at these radii. At small radii, the BCG mass density is approximately constant (Equation \ref{eqn:dpie}). From this and Equation \ref{eqn:eg_dm_reversed}, we find that
\begin{equation}
g_{\rm DM,EG}M_{\rm DM,EG} \approx \frac{2}{3}a_0M_{\rm B}
\end{equation}
For all four clusters we find that $g_{\rm DM,EG} \sim \frac{2}{3}a_0$ at small radii, indeed implying that $M_{\rm DM,EG} \sim M_{\rm B}$. 

We may also predict the apparent DM profile using the covariant version of EG from \citet{hossenfelder2017}, applying Equation \ref{eqn:ceg_dm_reversed} in the same way as Equation \ref{eqn:eg_dm_reversed}. We find (purple curves in Figure \ref{fig:eg1_mass}) that CEG only very slightly affects the predictions from EG, essentially translating the profile down by a factor $\sim 0.5$. CEG matches the apparent DM mass more precisely than the standard EG equation within $r \simlt 10$~kpc, but otherwise underpredicts at larger radii. 

The red curves in Figure \ref{fig:eg1_mass} also show the predicted apparent DM mass if the BCG were not present. Outside of $r \sim 100$~kpc, where the baryon mass is definitively dominated by the hot plasma, these curves agree with the green curves, but in the inner regions, they diverge significantly. 

This is shown more clearly in Figure \ref{fig:eg1_mass_ratio}, which shows the ratio of the observed DM mass to the predicted DM mass from EG with and without the BCG component. In agreement with previous results \citep{ettori2017,ettori2019}, the EG and the observations are in close agreement at $r \sim r_{500}$, but the actual DM mass is $\sim$2-4$\times$ higher at $r \sim 100$~kpc. The agreement with the observed DM is better if the BCG is included in the calculation (green curves) than if it is not (red curves). Given that CEG predicted a smaller apparent DM profile overall than EG, the same ratio is shown to have a greater discrepancy near $r \sim 100$~kpc than EG (purple curves). 

Figure \ref{fig:eg1_b_fraction} shows the baryon mass fraction $f_{\rm B} = M_{\rm B}(<r)/M_{\rm tot}(<r)$ profiles as a function of radius, both observed (orange curves) and predicted by EG for method EG1 (green curves). The observed baryon mass fraction begins at $\sim{0.6}$ in the cluster cores, where it is dominated by the stellar BCG component, and drops to $\sim{0.1}$ at $r\sim{100}$~kpc. Near this radius, where the baryons are now dominated by the hot plasma, $f_{\rm B}$ rises from $\sim{0.1}$ to nearly the cosmic baryon fraction $\Omega_{\rm b}/\Omega_{\rm m}$ at radii $r \simgt r_{500}$. In contrast, EG (green curves) underpredicts the baryon mass fraction slightly near the core, then decreases gradually, overpredicting from $r\sim{10-1000}$~kpc, where it intersects with the observed baryon fraction profile at $r\sim{1000}$~kpc and continues to decrease. If the BCG is not included in the calculation (red curves), the baryon fraction is nearly constant at $f_{\rm B} \sim 0.2$ over most of the radial range, and then begins to decrease near $r \sim 1000$~kpc. CEG predicts a larger baryon fraction with similar radial behavior than EG (purple curves), consistent with its lower prediction for the apparent DM profile. In any case, EG models predict that the baryon fraction should decrease with radius at large radii, which is inconsistent with the observations. 

As noted, Figures \ref{fig:eg1_mass}, \ref{fig:eg1_mass_ratio}, and \ref{fig:eg1_b_fraction} all show that the predictions from EG match the observations near $r_{500} \sim 1$~Mpc for our four clusters. The same agreement was also seen in different systems in \citet{ettori2017}, \citet{ettori2019}, and \citet{halenka2018}. Noting that for our clusters the two terms $M_B$ and $4\pi{r^3}\rho_B$ on the right-hand side of Equation \ref{eqn:eg_dm_reversed} are roughly equal at $r \sim r_{500}$, we can write:
\begin{equation}
M_{500}^2 \sim \frac{a_0}{3G}r_{500}^2M_{\rm B}(r_{500})
\end{equation}
Using the definition $M_{500} = 4\pi(500\rho_{\rm crit})r_{500}^3/3$, where $\rho_{\rm crit}$ is the critical density of the universe, and noting that $f_B \sim 0.15$ at $r_{500}$ for clusters of mass $M_{500} \sim 0.3-1.3 \times 10^{15} M_\odot$ such as ours \citep{vikhlinin2009}, we derive:
\begin{equation}
r_{500} \sim \frac{cf_B}{750H_0} \sim 0.9~{\rm Mpc}
\end{equation}
which is indeed fairly close to the values of $r_{500}$ for these clusters. Since this expression is comprised entirely of cosmological parameters and physical constants, it is not clear if it is anything more than a coincidence.

\subsection{Method EG2: Deriving Baryon Profiles from Total Mass Profiles}\label{sec:EG2}

Using method EG2 (Equation \ref{eqn:eg_dm_ode}), we take the total mass ($M_{\rm tot} = M_{\rm B}+M_{\rm DM}$) profile determined from gravitational lensing and determine the underlying baryonic profile which would generate the total gravitational acceleration in the EG framework, shown in Figure \ref{fig:eg2_mass}. At $r \simlt 10-30$~kpc, the EG prediction for the baryonic mass underpredicts the observed profile by a factor of $\sim 1.5-2.5$ (Figure \ref{fig:eg2_mass_ratio}), though the uncertainties in the innermost region are large. At larger radii ($r \sim 100-1000$~kpc), EG overpredicts the baryon mass by a factor of $\sim 3-4$. The baryon mass profile predicted by method CEG2 (lime curves) is typically $\sim$~30-50\% higher than that of EG2, and thus in greater discrepancy from the observed profile. 

Figure \ref{fig:eg2_b_fraction} shows the predicted baryon mass fraction profiles from method EG2 (brown curves), compared to the observed profiles (blue curves). Within the cores of the clusters, $r \simlt 10-30$~kpc, the baryon mass fraction is slightly underpredicted, as the predicted baryon profile is less than the observed BCG profile. The predicted baryon fraction from EG2 gradually decreases and greatly overpredicts the observed baryon fraction, which decreases faster before increasing around $r \sim 100$~kpc. These two profiles intersect near $r \simgt r_{500}$. The predicted baryon fraction profiles from CEG2 are higher than that from EG2 at radii less than $r_{500}$ with similar behavior. 

\begin{figure}
\centering
\includegraphics[width=0.45\textwidth]{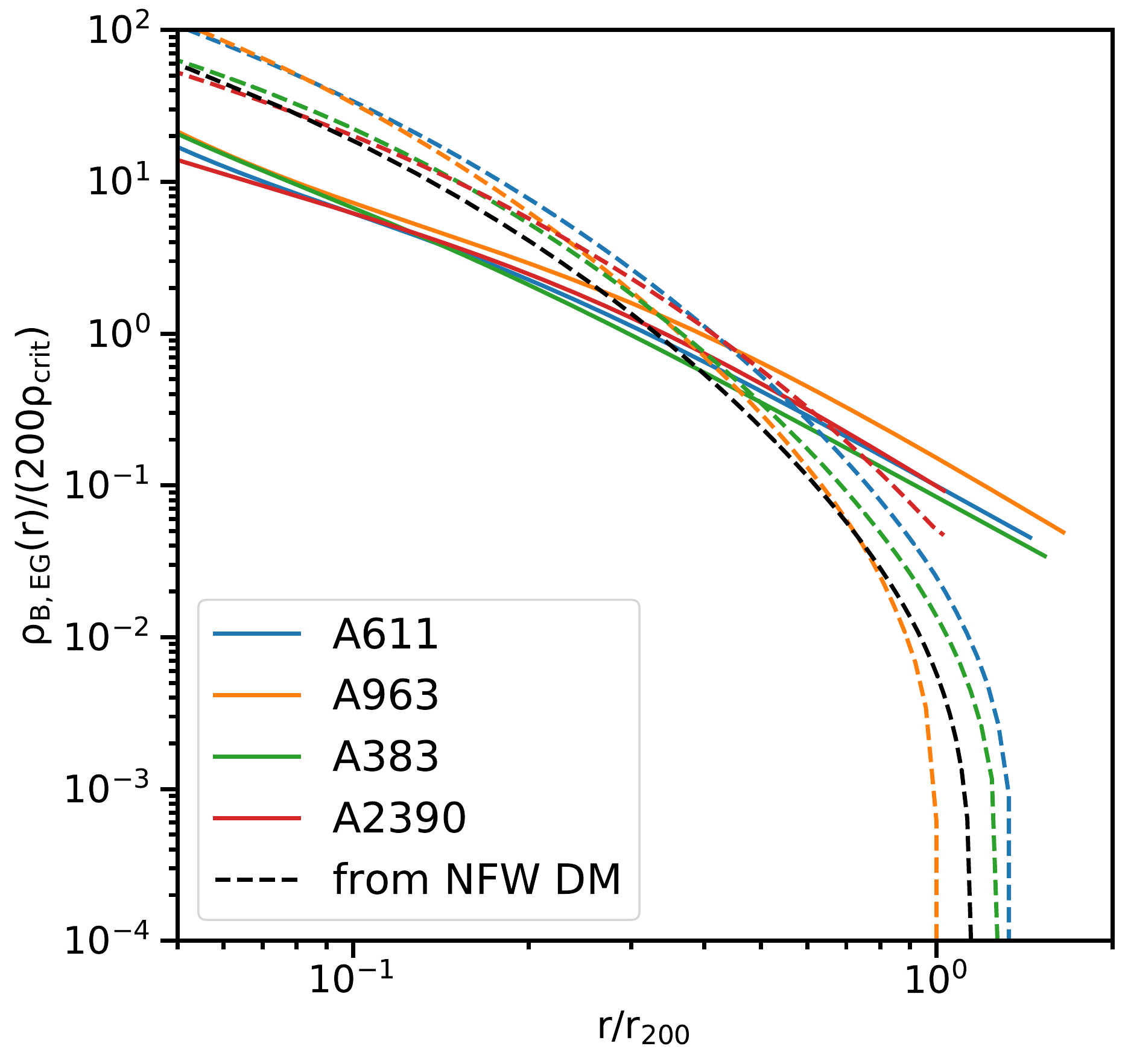}
\caption{The baryon density profiles predicted by method EG2 at large radii (dashed lines, scaled by their $r_{200}$) for our four clusters, scaled by 200 times the critical density of the universe. The observed baryon densities are shown by the solid lines. The predicted baryon density for a cluster of similar mass with a NFW DM density profile with concentration parameter $c_{200}$ = 4 is shown by the dashed black line.\label{fig:neg_dens}}
\end{figure}

Figure \ref{fig:neg_dens} shows the predicted baryon profiles for all four clusters using method EG2, and the observed baryon density profiles. We note that at large radii the predicted baryonic matter density for all four clusters drops below zero and becomes negative--a clearly unphysical result (with the exception of A2390, which may presumably have the same behavior at a slightly larger radius than we have available). The same effect is present whether we use the original EG equation or the CEG equation. The same result can be obtained for a generic cluster by assuming $M_{\rm DM,EG}$ is of NFW form in Equation \ref{eqn:eg_dm}, shown by the black dashed profile in Figure \ref{fig:neg_dens}. We find that this appearance of negative baryon densities is common to apparent DM density profiles which have $\rho_{\rm DM,EG} \sim {r}^{-3}$ (NFW) or steeper at large radii. This is significant, as X-ray and gravitational lensing analyses have demonstrated that the mass profiles of galaxy clusters are well-fit by NFW profiles \citep{broadhurst2005,newman2013a,newman2013b,umetsu2017,ettori2019}.

\section{Discussion and Conclusions}\label{sec:conc}

In this work, we have used observed baryonic and total mass profiles of four massive and relaxed galaxy clusters to test if the Emergent Gravity theory of \citet{verlinde2017} is a viable alternative to cold dark matter. We extend the work of \citet{ettori2017}, \citet{ettori2019}, and \citet{halenka2018} by considering the contribution of the BCG to the mass and density profiles of each galaxy cluster, and by also performing the same set of calculations for the Covariant Emergent Gravity formulation of \citet{hossenfelder2017}. To evaluate the theory we first use the baryon profile to predict the apparent DM profile, and we also separately use the total apparent mass profile to predict the baryon profile which would generate it. Intriguingly, we find that the addition of the BCG improves the agreement of the observations with the predictions of EG in the core regions of clusters relative to these earlier results.

However, we also find that Emergent Gravity in its current form is still not a viable alternative to the standard dark matter model on galaxy cluster scales, as it does not make predictions which are consistent with the observations of galaxy clusters over most of their radial extent. In particular, at radii $r \simgt 10$~kpc, and especially near $r \sim 100$~kpc, EG underpredicts the apparent DM mass assuming the underyling baryon profile, and it overpredicts the baryon mass assuming the total apparent mass profile. EG also predicts that the baryon mass fraction will decrease with radius from the center of a cluster to the outskirts, but in the outer regions of clusters the baryon mass fraction increases with radius. For NFW-like DM mass profiles which are observed for clusters, EG predicts the baryon density will drop to zero and become negative. There is good agreement between the predictions of EG and the observations at $r \sim r_{500}$ for all four of the clusters we studied, but this appears to be a numerical coincidence. In sum, the theory of EG as it is presently constructed does not appear to make predictions for the apparent DM and baryon mass densities of galaxy clusters which are in agreement with observation. This is in line with the previous investigations using EG in the context of galaxy clusters \citep{ettori2017,ettori2019,halenka2018}, as well as similar attempts to explain the total gravitational acceleration of clusters using MOND \citep{aguirre2001,sanders2003}, which has a similar functional form. 

This work made a number of simplifying assumptions which limit its applicability, in particular those of spherical symmetry and hydrostatic equilibrium. Though the clusters we analyzed here are reasonably ``relaxed'', even relaxed clusters show evidence for gas motions \citep{hitomi2016,hitomi2018}. In general, clusters are also not spherically symmetric but possess a degree of ellipticity \citep[e.g.][]{flin1984,evans2009,oguri2010,lau2012,shin2018}. Addressing these issues would require a more general analysis which is not restricted to spherical systems in hydrostatic equilibrium, requiring a more general framework such as CEG \citep{hossenfelder2017}. 

Another source of uncertainty is that for the clusters we examined, the X-ray data at large radii ($r \simgt r_{500}$) do not constrain well the form of the baryon density profile in this region. Other authors have performed similar fits to the X-ray data for these clusters, obtaining slightly different results for the gas densities at large radii. To test the effect of these variations, we carried out our calculations using the fitted profiles for our four clusters using parameter fits from \citet[][for A383 and A2390]{vikhlinin2006} and \citet[][for A611, A963, and A2390]{giles2017}. We find minor differences in the computed profiles at $\simlt 10\%$ levels, mostly at $r \simgt r_{500}$, which do not affect our conclusions. 

Though it appears that particle DM is required to explain the gravitational potentials of galaxy clusters as inferred from gravitational lensing, member galaxy motions, and X-ray observations \citep[in addition to dissociative cluster mergers with large separations between lensing and X-ray peaks such as the Bullet Cluster;][]{clowe2006}, the agreement of EG in the BCG-dominated inner regions of our relaxed clusters is intriguing. It is also in line with the success of the EG model on galactic scales from other investigations, as well as the success of MOND-like models in general on these scales. Recently, a model for DM has been advanced with two phases: a particle-like phase which represents the standard CDM model and applies on galaxy cluster scales, and a superfluid-like phase which sets in on galaxy scales \citep{bere2015,khoury2016,hodson2017b,bere2018}. This latter phase includes a phonon-mediated force between the DM and baryons with similar properties to MOND and EG \citep{hossenfelder2018}. Such a model may explain the improved agreement between the predictions of EG with the observations within $r \simlt 10$~kpc in all four of the clusters studied in this work as shown in Figure \ref{fig:eg1_mass_ratio}. Though only tested so far in simple, spherically-symmetric models such as ours, this physics could be incorporated into simulations of galaxy formation and evolution. 

\acknowledgments
The authors thank Andrew Newman for supplying us with the profiles for the BCG and the total mass density from \citet{newman2013b}, Pasquale Mazzotta for supplying us with the gas density profiles from \citet{martino2014}, and Alexey Vikhlinin for providing the gas density data for A383 from \citet{vikhlinin2006} for comparison. We also thank Sabine Hossenfelder, Vittorio Ghirardini, and Pasquale Mazzotta for useful comments. JAZ acknowledges support from NASA contract NAS8-03060 with the {\it Chandra} X-ray center.

\end{document}

%% file: clusters.tex
\begin{table*}[]
\centering
\caption{Cluster Properties}
\begin{tabular}{llllll}
\hline
\hline
Name & Redshift & $M_{200}~(10^{15}~M_\odot)$ & $r_{200}$ (Mpc) & $M_{500}~(10^{15}~M_\odot)$ & $r_{500}$ (Mpc) \\
\hline
A383 & 0.190 & 0.41 & 1.29 & 0.30 & 0.85 \\
A611 & 0.288 & 0.62 & 1.36 & 0.45 & 0.90 \\
A963 & 0.206 & 0.34 & 1.20 & 0.26 & 0.81 \\
A2390 & 0.229 & 1.48 & 1.91 & 0.99 & 1.23 \\
\hline
\end{tabular}
\label{tab:clusters}
\end{table*}